\documentclass[iop]{emulateapj}

\newcommand{\swift}{{\it Swift}}
\newcommand{\ngc}{NGC\, 4151}
\newcommand{\et}{et al.\,}

\newcommand{\edt}[1]{{#1}}
\newcommand{\edit}[1]{{#1}}

\shorttitle{Swift monitoring of NGC 4151}
\shortauthors{Edelson \et}

\begin{document}

\title{Swift monitoring of NGC 4151: Evidence for a Second X-ray/UV Reprocessing}

\author{
R.~Edelson\altaffilmark{1},
J.~Gelbord\altaffilmark{2},
E.~Cackett\altaffilmark{3},
\edit{S.~Connolly\altaffilmark{4},}
C.~Done\altaffilmark{5},
M.~Fausnaugh\altaffilmark{6},
E.~Gardner\altaffilmark{5},
N.~Gehrels\altaffilmark{7,35},
M.~Goad\altaffilmark{8},
K.~Horne\altaffilmark{9},
I.~M$^{\rm c}$Hardy\altaffilmark{4},
B.~M.~Peterson\altaffilmark{6,10,11},
S.~Vaughan\altaffilmark{8},
M.~Vestergaard\altaffilmark{12,13},
A.~Breeveld\altaffilmark{14},
A.~J.~Barth\altaffilmark{15},
M.~Bentz\altaffilmark{16},
M.~Bottorff\altaffilmark{17},
W.~N.~Brandt\altaffilmark{18,19,20},
S.~M.~Crawford\altaffilmark{21},
E.~Dalla~Bont\`{a}\altaffilmark{22,23},
\edit{D.~Emmanoulopoulos\altaffilmark{4},}
P.~Evans\altaffilmark{8},
R.~Figuera~Jaimes\altaffilmark{9},
A.~V.~Filippenko\altaffilmark{24},
G.~Ferland\altaffilmark{25},
D.~Grupe\altaffilmark{26},
M.~Joner\altaffilmark{27},
J.~Kennea\altaffilmark{18},
K.~T.~Korista\altaffilmark{28},
H.~A.~Krimm\altaffilmark{7,29},
G.~Kriss\altaffilmark{10},
D.~C.~Leonard\altaffilmark{30},
S.~Mathur\altaffilmark{5},
H.~Netzer\altaffilmark{31},
J.~Nousek\altaffilmark{18},
K.~Page\altaffilmark{8},
E.~Romero-Colmenero\altaffilmark{21,32},
M.~Siegel\altaffilmark{15},
D.~A.~Starkey\altaffilmark{9},
T.~Treu\altaffilmark{33},
H.~A.~Vogler\altaffilmark{15},
H.~Winkler\altaffilmark{34},
and 
W.~Zheng\altaffilmark{24}
}
\altaffiltext	{1}{ University of Maryland, Department of Astronomy, College Park, MD 20742-2421, USA}
\altaffiltext	{2}{Spectral Sciences Inc., 4 Fourth Ave., Burlington, MA 01803, USA}
\altaffiltext	{3}{Department of Physics and Astronomy, Wayne State University, 666 W. Hancock St, Detroit, MI 48201, USA}
\altaffiltext	{4}{University of Southampton, Highfield, Southampton, SO17 1BJ, UK}
\altaffiltext	{5}{University of Durham, Center for Extragalactic Astronomy, Department of Physics, South Rd., Durham, DH1 3LE, UK}
\altaffiltext	{6}{The Ohio State University, Department of Astronomy, 140 W 18th Ave, Columbus, OH 43210, USA}
\altaffiltext	{7}{Astrophysics Science Division, NASA Goddard Space Flight Center, Greenbelt, MD 20771, USA}
\altaffiltext	{8}{University of Leicester, Department of Physics and Astronomy, Leicester, LE1 7RH, UK}
\altaffiltext	{9}{SUPA Physics and Astronomy, University of St. Andrews, Fife, KY16 9SS Scotland, UK}
\altaffiltext	{10}{Space Telescope Science Institute, 3700 San Martin Drive, Baltimore, MD 21218, USA}
\altaffiltext	{11}{Center for Cosmology and AstroParticle Physics, The Ohio State University, 191 West Woodruff Avenue, Columbus, OH, 43210, USA}
\altaffiltext	{12}{Dark Cosmology Centre, Niels Bohr Institute, University of Copenhagen, Juliane Maries Vej 30, DK-2100 Copenhagen, Denmark}
\altaffiltext	{13}{Steward Observatory, University of Arizona, 933 North Cherry Avenue, Tucson, AZ 85721, USA}
\altaffiltext	{14}{Mullard Space Science Laboratory, University College London, Holmbury St. Mary, Dorking, Surrey RH5 6NT, UK}
\altaffiltext	{15}{Department of Physics and Astronomy, 4129 Frederick Reines Hall, University of California, Irvine, CA 92697, USA}
\altaffiltext	{16}{Department of Physics and Astronomy, Georgia State University, 25 Park Place, Suite 605, Atlanta, GA 30303, USA}
\altaffiltext	{17}{Physics Department, Southwestern University, Georgetown, TX 78626, USA}
\altaffiltext	{18}{Department of Astronomy and Astrophysics, Eberly College of Science, 525 Davey Laboratory, The Pennsylvania State University, University Park, PA 16802, USA}
\altaffiltext	{19}{Institute for Gravitation and the Cosmos, The Pennsylvania State University, University Park, PA 16802, USA}
\altaffiltext	{20}{Department of Physics, 104 Davey Laboratory, The Pennsylvania State University, University Park, PA 16802, USA}
\altaffiltext	{21}{South African Astronomical Observatory, P.O. Box 9, Observatory 7935, Cape Town, South Africa}
\altaffiltext	{22}{Dipartimento di Fisica e Astronomia ``G. Galilei,'' Universit\`{a} di Padova, Vicolo dell'Osservatorio 3, I-35122 Padova, Italy}
\altaffiltext	{23}{INAF-Osservatorio Astronomico di Padova, Vicolo dell'Osservatorio 5 I-35122, Padova, Italy}
\altaffiltext	{24}{Department of Astronomy, University of California, Berkeley, CA 94720-3411, USA}
\altaffiltext	{25}{Department of Physics and Astronomy, University of Kentucky, Lexington KY 40506, USA}
\altaffiltext	{26}{Space Science Center, Morehead State University, 235 Martindale Dr., Morehead, KY 40351, USA}
\altaffiltext	{27}{Department of Physics and Astronomy, N283 ESC, Brigham Young University, Provo, UT 84602, USA}
\altaffiltext	{28}{Department of Physics, Western Michigan University, 1120 Everett Tower, Kalamazoo, MI 49008, USA}
\altaffiltext	{29}{Universities Space Research Association, 7178 Columbia Gateway Drive, Columbia, MD 21046, USA}
\altaffiltext	{30}{Department of Astronomy, San Diego State University, San Diego, CA 92182, USA}
\altaffiltext	{31}{School of Physics and Astronomy, Raymond and Beverly Sackler Faculty of Exact Sciences, Tel Aviv University, Tel Aviv 69978, Israel}
\altaffiltext	{32}{Southern African Large Telescope Foundation, P.O. Box 9, Observatory 7935, Cape Town, South Africa}
\altaffiltext	{33}{Department of Physics and Astronomy, University of California, Los Angeles, CA 90095-1547, USA}
\altaffiltext	{34}{Department of Physics, University of Johannesburg, PO Box 524, 2006 Auckland Park, South Africa}
\altaffiltext	{35}{Deceased}

\begin{abstract}
\swift\ monitoring of \ngc\ with $\sim$6~hr sampling over a total of 69 days in early 2016 is used to construct light curves covering five bands in the X-rays (0.3--50~keV) and six in the ultraviolet (UV)/optical (1900--5500~\AA).
The three hardest X-ray bands (\edit{$>$2.5~keV}) are all strongly correlated with no measurable interband lag, while the two softer bands show lower variability and weaker correlations.
The UV/optical bands are significantly correlated with the X-rays, lagging $\sim$3--4~days behind the hard X-rays.
The variability within the UV/optical bands is also strongly correlated, with the UV appearing to lead the optical by $\sim$\edit{0.5}-1~day.
This combination of $\gtrsim$3~day lags between the X-rays and UV and $\lesssim$1~day lags within the UV/optical appears to rule out the ``lamp-post'' reprocessing model in which a hot, X-ray emitting corona directly illuminates the accretion disk, which then reprocesses the energy in the UV/optical.
Instead, these results appear consistent with the Gardner \& Done picture in which two separate reprocessings occur:
first, emission from the corona illuminates an extreme-UV-emitting toroidal component that shields the disk from the corona;
this then heats the extreme-UV component, which illuminates the disk and drives its variability.
\end{abstract}


\section{Introduction}
\label{section:intro}

Although the quantity and quality of observational data on active galactic nuclei (AGN) has vastly improved over the past few decades, the standard model of the physical structure of the central engine has remained largely unchallenged.
The fundamental picture of \edit{an accretion disk} surrounding a supermassive black hole (SMBH) was first proposed by \cite{Lynden-Bell69}.
The model of an optically thick, geometrically thin accretion disk was first proposed by \cite{Shakura73} in the context of stellar-mass black holes.
\cite{Galeev79} added magnetic reconnection in a corona above the disk in order to explain the observed hard X-ray emission from AGN. 
This predicts that the corona can directly illuminate and heat the outer disk  (e.g., \citealt{Frank02}), leading to the so-called  ``lamp-post'' or ``reprocessing'' model.
Note that in this paper the use of the term ``lamp-post'' does not require that the X-ray source must be a point source;
instead, we only require that it is small relative to the  ultraviolet and optical (UV/optical) emitting disk, extending above and below the disk so that it directly illuminates the disk.

A clear prediction of this model is that flux variations in the X-ray emitting corona will be seen in the UV/optical emission from the disk.
Measurement of the interband X-ray/UV temporal lag and smoothing can then be used to estimate the size and structure of the disk.
This technique, known as reverberation mapping (RM; \citealt{Blandford82}; \citealt{Peterson93}; \citealt{Peterson14}), has been used for decades in a different context to constrain the size and physical characteristics of the broad emission-line region (BLR; \citealt{Peterson97}).
This model predicts a clear relation between lag ($\tau$) and wavelength ($\lambda$) as the variations from the smaller, hotter inner disk are expected to precede those from the larger, cooler outer disk regions, scaling as $ \tau \propto \lambda^{4/3} $ (e.g., \citealt{Cackett07}).

Application of RM to the corona/disk system has been more difficult than to the BLR because the sizes (and thus the lags) are much smaller ($\lesssim$1 day).
Nonetheless, this ``accretion disk RM'' approach has been repeatedly attempted because of the potential large reward: information on the size and structure of the central engines of AGN that cannot be probed by any other method except gravitational lensing in rare cases (e.g., \citealt{Morgan10}).
Most early disk RM experiments yielded inconclusive results.
For instance, an early campaign on \ngc\ built around the {\it International Ultraviolet Explorer} found a hint of the shorter-wavelength UV leading longer wavelengths, though the measured lag was not significantly different from zero (\citealt{Crenshaw96}; \citealt{Edelson96}).
Further efforts to implement disk RM by correlating X-ray light curves gathered with space-based observatories with optical light curves typically from ground-based observatories (e.g., \citealt{Wanders97}; \citealt{Collier98}; \citealt{Nandra98};  \citealt{Collier99}; \citealt{Collier01}; \citealt{Suganuma06}; \citealt{Arevalo08}; \citealt{Arevalo09}; \citealt{Breedt09}; \citealt{Breedt10}; \citealt{Cameron12}; \citealt{Gliozzi13}) have often yielded suggestions of interband lags in the expected direction, but the results were never statistically significant ($>3\sigma$).
Likewise, ground-based optical monitoring also yielded indications that the shorter wavelengths led the longer wavelengths (e.g., \citealt{Sergeev05}; \citealt{Cackett07}), but again not at a statistically significant level.

Recent observations have been able to produce more solid results by taking advantage of the unique capabilities of the \swift\ satellite, in particular, its ability to sample at high cadence across the X-ray/UV/optical regime needed to perform this experiment.
\cite{Shappee14} and \cite{McHardy14} find clear evidence of the UV leading the optical in NGC~2617 and NGC~5548, respectively.
The clearest previous measurement of interband lags was seen in a very large ($\sim$300 observations) \swift/{\it HST}/ground-based campaign on NGC~5548 (\citealt{Edelson15}; \citealt{Fausnaugh16}).
A recent archival \swift\ survey by \cite{Buisson16} reports evidence that X-ray variations lead the UV in several AGN, but also shows that detailed disk RM requires long-duration, high-cadence campaigns with multi-filter \swift\ UltraViolet/Optical Telescope (UVOT; \citealt{Roming05}) data similar to what was done for NGC~5548.

This paper reports the results of intensive \swift\ monitoring of \ngc, with particularly detailed coverage in the X-ray regime, allowing us to measure temporal correlations and time lags between bands spanning an unprecedented wavelength range out to 50~keV.
These results contradict the standard reprocessing model because the X-ray/UV lags are observed to be much longer than those within the UV/optical.
This indicates that the arrangement of the emission components cannot be as simple as an X-ray corona that directly illuminates and drives a UV/optical-emitting accretion disk.
Instead, the interband lags are consistent with the picture proposed by \cite{Gardner17}, which posits the existence of an energetically important emission component that peaks in the unobservable extreme-ultraviolet (EUV).
While the peak of this putative component in the EUV cannot be directly observed, the ``soft excess'' seen in the X-rays and the ``big blue bump'' seen in the UV/optical could be interpreted as its high/low-frequency tails.
The observed interband lags appear to be consistent with such an EUV component acting as an additional reprocessor that is illuminated and heated by the X-ray corona and then in turn illuminates and drives the variability in the accretion disk.

This paper is organized as follows.
Section~2 summarizes the observations and data reduction,  
Section~3 presents the timing analysis, 
Section~4 discusses the challenges these results present for the standard reprocessing model and how the addition of a EUV component may solve these problems,
and Section~5 gives some brief concluding remarks.

\section{Observations and data reduction}
\label{section:data}

\subsection{Target}
\label{section:target}

The target of this experiment, the Seyfert~1.5 galaxy \ngc\ (redshift $ z = 0.00332 $, \citealt{deVaucouleurs91}; distance $ D \approx 19$~Mpc; \citealt{Hoenig14}), is typically the brightest Seyfert~1 galaxy in the sky in the X-ray/UV/optical wavelength range accessible to \swift.
For instance, the \swift\ Burst Alert Telescope (BAT) catalog \citep{Krimm13} indicates that \ngc\ is twice as bright in the 15--50~keV band as the next-brightest Type 1 AGN.
\ngc\ is one of the Seyfert~1 galaxies in the original identification paper on these objects \citep{Seyfert43} and is often considered to be an archetype of the class \citep{Ulrich00}. 
It is well known to be strongly variable across the wavelength range accessible to \swift\ \citep{Edelson96}, making it an ideal monitoring target.

The bolometric luminosity of \ngc\ is $L_{\rm bol} \approx 5 \times 10^{43}$ erg s$^{-1}$ \citep{Woo02}.
The central black hole mass has been measured by RM to be $ M_{\rm BH} \approx 43.57^{+0.45}_{-0.37} \times 10^{7}\,{\rm M}_{\odot}$ (\citealt{Bentz06}, updated with the calibration of \citealt{Grier13}), by gas dynamics to be
$3.0^{+0.75}_{-2.2} \times 10^7 \,{\rm M}_{\odot}$ \citep{Hicks08}, and by stellar dynamics to be $(3.76\pm1.11)\times 10^7\,{\rm M}_{\odot}$ \citep{Onken14}.
\edit{We adopt a value of $M_{\rm BH} \sim 4 \times 10^{7}\,{\rm M}_{\odot}$ in this paper.}

\ngc\ has been particularly well-studied in the X-rays. 
The soft X-rays are only weakly variable because that band is dominated by extended line emission (e.g., \citealt{Zdziarski02}), but at higher energies (above $\sim$2~keV) the flux is strongly variable and thought to be coming from the corona.
{\it NuSTAR/Suzaku} spectroscopy is consistent with reflection from the inner disk in \ngc\ \citep{Keck15}.
X-ray time lags also show Fe K$\alpha$ reverberation in this object that would require reflection from the inner disk (\citealt{Zoghbi12}; \citealt{Cackett14}).

\subsection{Observations}
\label{section:obs}

During 2016 February 20 through April 29, \swift\ executed an intensive monitoring campaign on \ngc\ consisting of 319 separate visits of at least 120~s, an average of nearly 5 visits per day.
These observations are summarized in Table~1.
Start and stop times for \swift\ observations are originally recorded in Mission Elapsed Time (seconds since the start of 2001) and corrected for the drift of the on-board \swift\ clock and leap-seconds.
These times were converted to Modified Julian Date (MJD), the standard for this observing campaign.

\begin{deluxetable}{lccccr}
\tablecaption{Monitoring Information \label{table1}}
\tablewidth{0pt}
\tablecolumns{6}
\tablehead{
\colhead{(1)} & \colhead{(2)} & \colhead{(3)} & \colhead{(4)} & \colhead{(5)} & \colhead{(6)} \cr
 \colhead{} & \colhead{Central} & \colhead{Wavelength/} & 
  \colhead{Number} & \colhead{Sampling}  & $F_{\rm var}$ \cr
 \colhead{Band} & \colhead{$\lambda$ (\AA)} & \colhead{Energy Range} 	& \colhead{of Points} & \colhead{Rate (day)}  & (\%) \cr}
\startdata
 BAT        & 0.45 & 15--50 keV      &  69 & 1.00 & 18.6 \cr
 X4         &  1.8 & 5--10 keV       & 319 & 0.22 & 34.6 \cr
 X3         &  3.5 & 2.5--5 keV      & 319 & 0.22 & 41.6 \cr
 X2         &    7 & 1.25--2.5 keV   & 319 & 0.22 & 17.4 \cr
 X1         &   20 & 0.3--1.25 keV   & 319 & 0.22 &  9.1 \cr
 {\it uvw2} & 1928 & 1650--2250 \AA\ & 254 & 0.28 &  6.1 \cr
 {\it uvm2} & 2246 & 2000--2500 \AA\ & 252 & 0.23 &  5.7 \cr
 {\it uvw1} & 2600 & 2250--2950 \AA\ & 273 & 0.26 &  5.4 \cr
 {\it u}    & 3465 & 3050--3900 \AA\ & 276 & 0.22 &  6.0 \cr
 {\it b}    & 4392 & 3900--4900 \AA\ & 319 & 0.22 &  3.9 \cr
 {\it v}    & 5468 & 5050--5800 \AA\ & 310 & 0.23 &  2.4 \cr
\enddata
\tablecomments{Column 1: observing band name.
Column 2: central wavelength of that band.
Column 3: wavelength/energy range covered by each band.  For the five X-ray bands (top) the range is given in keV.  For the six UVOT bands, the FWHM wavelength range is given in \AA, estimated from \cite{Poole08}.
Column 4: total number of good data points in that band.
Column 5: mean sampling interval in that band.
Column 6: fractional variability amplitude, $F_\mathrm{var}$, as defined by \cite{Vaughan03}, not corrected for the constant galaxy contribution.}
\end{deluxetable}

\swift\ observations with the UVOT were made in mode 0x037a, which allows for hardware windowing in the four longest-wavelength bands.
This was done because this source is too bright to be observed in a standard, non-windowed mode. 
Observations with the X-Ray Telescope (XRT; \citealt{Burrows05}) were made in Photon Counting (PC) mode, except for the last seven, which were made in Windowed Timing (WT) mode \citep{Hill04}.
The impacts of these observing modes on the data quality and other details are discussed in the following subsections.

These \swift\ observations were coordinated with intensive monitoring with numerous ground-based telescopes including the Las Cumbres Observatory Global Telescope (LCOGT) network and the Liverpool Telescope at La Palma.
Those data will be presented in subsequent papers (K. Horne \et in preparation; M. Goad \et in preparation).

\subsection{UVOT Data Reduction}
\label{section:uvot}

\edit{The UVOT data were taken in} a six-filter, blue-weighted mode
\edit{in which the four longest-wavelength filters}
({\it uvw1, u, b}, and {\it v})  
\edit{are observed using}
5\arcsec $\times$ 5\arcsec\ 
\edit{hardware} windows.
\edit{These reduce} the frame time from 11 to 3.6~ms, 
\edit{thereby mitigating} the effect of pile-up (coincidence losses) in this bright source.
The four hardware window observations are preceded by short (10~s) full-field exposures, 
\edit{but these are not used because the coincidence losses were found
  to be too large to be corrected reliably.}
This mode 
\edit{also splits the {\it uvm2} data into two exposures when the
  time exceeds 300~s; such exposure pairs that survive screening are
  co-added before final analysis.}

All UVOT data were reprocessed for uniformity, applying standard {\tt FTOOLS} utilities (\citealt{Blackburn95}; from version 6.19 of {\tt HEASOFT}\footnote{http://heasarc.gsfc.nasa.gov/ftools/}).  
The astrometry of each field was refined using the AGN and up to 25 isolated field stars drawn from the {\it HST} GSC 2.3.2 \citep{Lasker08} and Tycho-2 \citep{Hog00} catalogs, yielding residual offsets that were typically $\sim$0.3\arcsec.
Fluxes were measured using a 5\arcsec\ \edit{radius} circular aperture, and concentric 40--90\arcsec\ \edit{radius} annuli were used to measure the sky background level.
The final values include corrections for aperture losses, coincidence losses, large-scale variations in the detector sensitivity across the image plane, and declining sensitivity of the instrument over time.
After reprocessing, \edit{25 exposures} were screened out to eliminate
observations affected by tracking errors or with exposure times
shorter than 20~s.

We use a non-default setting 
\edit{when accounting for systematic errors in the aperture correction
  arising from variations in the UVOT point-spread function.
  The tool {\tt UVOTAPERCORR} normally adds a filter-dependent
  uncertainty of 1.85--2.15\% to measurements made with a
  5\arcsec\ aperture.}
However, when measuring
\edit{fluxes of}
\ngc\ and the field stars with the highest signal-to-noise ratios (S/Ns), we found the resulting error estimates to be inconsistent with Gaussian statistics.
This result is not surprising given that the {\tt UVOTAPERCORR} documentation 
\edit{notes that the appropriate size of the systematic error}
is not well established.
We empirically examined a range of 
\edit{systematic error estimates by adjusting the parameter FWHMSIG}
and found that halving this parameter (to 7.5) yielded distributions much more consistent with Gaussian.
For instance, 
\edit{in the case of the UV-brightest star in the field (BD+40 2507),}
88.7\% of the {\it uvm2} and 76.8\% of the {\it uvw2}
measurements fall within $\pm 1 \sigma$ of the mean
\edit{when using the default FWHMSIG setting, whereas}
these percentages are 72.8\% and 64.1\% 
when FWHMSIG = 7.5.
\edit{By adopting this setting, the flux uncertainties reported here include}
filter-dependent systematic errors of 0.92--1.08\%.

\begin{figure}
\figurenum{1}
\begin{center}
 \includegraphics[width=3.9in,angle=270]{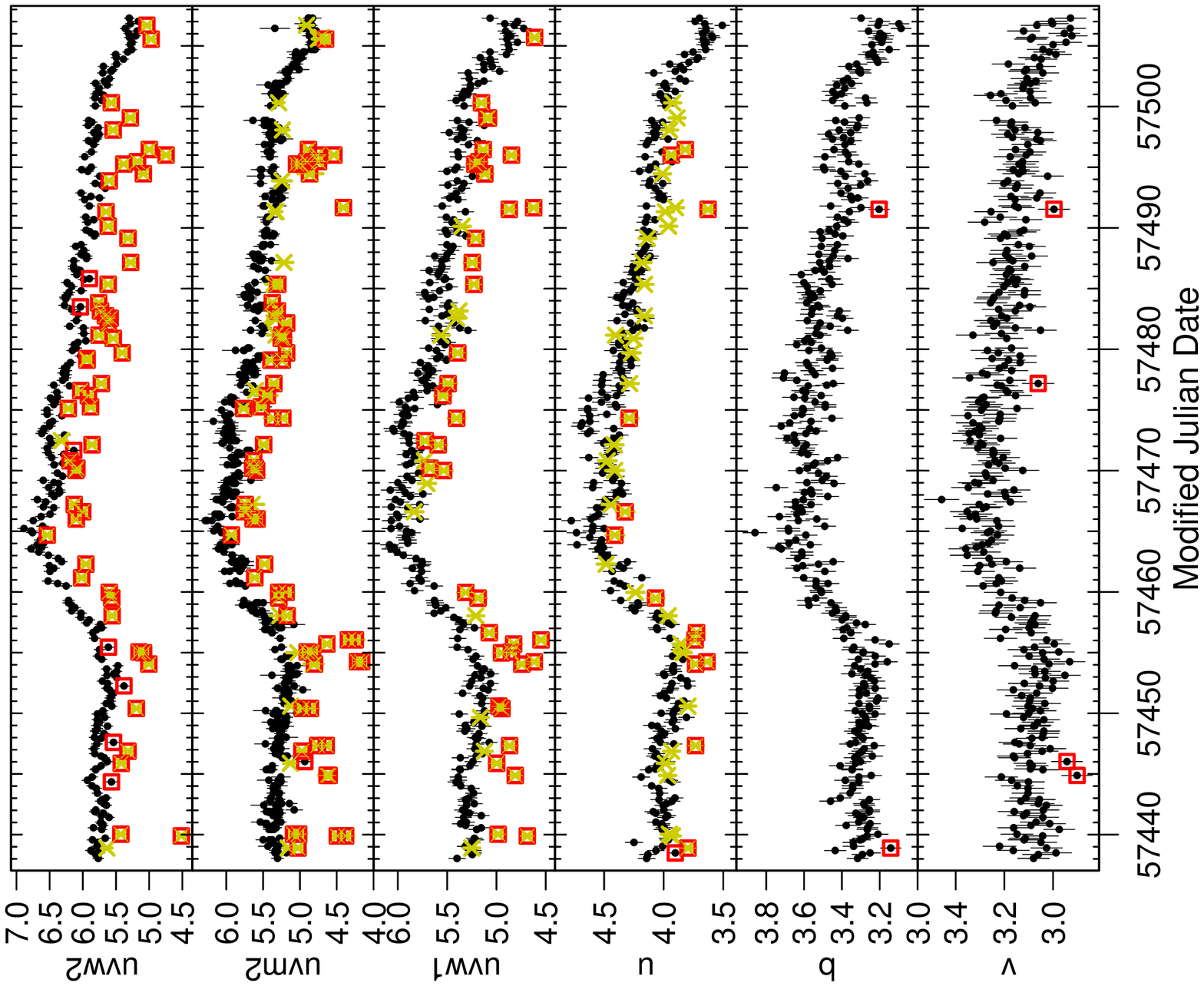} 
\caption{\ngc\ light curves before filtering.
Y-axis fluxes are given in units of $10^{-14}$ erg cm$^{-2}$ s$^{-1}$ \AA$^{-1}$.
Points flagged as dropouts are shown as red boxes.
These points are also shown in red in Figure~2, where they are used to define ``bad detector regions.''
Points in the four shortest-wavelength bands that fell in those bad regions (yellow in Figure 2) are also shown in yellow in this figure.
Finally, the black points are the remaining good data that passed our filtering.}
\label{fig:fig1}
\end{center}
\end{figure}

The resulting light curves, shown in Figure~\ref{fig:fig1}, exhibited occasional anomalously low points (``dropouts''), especially in the UV. 
Similar dropouts were seen in an earlier \swift\ study of NGC 5548 \citep{Edelson15} and 
\edit{were}
found to be clustered in the detector plane.
This may be due to localized regions of reduced sensitivity (\citealp{Breeveld16};
it should be noted that the deviant flux points 
\edit{are universally low, not symmetric about the light curve as
  would be expected from AGN variability).}
As such, we filtered discrepant points in the \ngc\ data in a fashion similar to that in the Appendix of \citeauthor{Edelson15}
This filtering consists of four steps:
(1) identify dropouts from the light curves;
(2) map the data onto the detector plane;
(3) define boxes \edit{to enclose clusters of bad data}; and
(4) use 
\edit{the set of boxes as a mask and screen out}
all data in these regions in the four shortest-wavelength bands.
\edit{The procedure used here differs from \cite{Edelson15}\ in that we
model the light curve by fitting}
a quadratic expression to each light curve in a sliding window of $\pm$2 days.

\edit{We define dropouts as points below the light curve with absolute
  deviations that exceed those of the highest positive outliers in the entire light curve.}
The number found in each band is given in Table~\ref{table2}.
\edit{Dropouts are most prevalent at short wavelengths, accounting for $>$10\% of measurements in the UV bands, nearly 5\% in {\it u}, and are barely found in {\it b} and {\it v}.}

\begin{deluxetable}{lccccc}
\tablecaption{UVOT Dropout Testing \label{table2}}
\tablecolumns{6}
\tablehead{
\colhead{(1)} & \colhead{(2)} & \colhead{(3)} & 
 \colhead{(4)} & \colhead{(5)} & \colhead{(6)} \cr
\colhead{Filter} & \colhead{Num} & \colhead{Dropouts} & 
 \colhead{Num}    & \colhead{Masked} & \colhead{Final} \cr
\colhead{}  & \colhead{Obs} & \colhead{} & \colhead{Masked} &  \colhead{Non-drop} & \colhead{Data}}
\startdata
{\it uvw2}	&	308	&	57	&	54	&	2	 &	254	\cr
{\it uvm2}	&	455	&	62	&	80	&	19 &	252\tablenotemark{a} \cr
{\it uvw1}	&	322	&	38	&	49	&	11 &	273	\cr
{\it u}	    &	320	&	14	&	44	&	31 &	276	\cr
{\it b}	    &	319	&	2	  &	0   &	0	 &	319	\cr
{\it v}	    &	310	&	4	  &	0   &	0  &	310 
\enddata
\tablenotetext{a}{\edit{After co-adding pairs of {\it uvm2} exposures.}}
\tablecomments{\edit{Columns are (1) the UVOT filter, followed by counts of
  (2) points in the light curve used for dropout testing, (3) exposures
  flagged as dropouts and used to define masks of suspect
  detector regions, (4) measurements made within these regions, (5)
  the subset of (4) not flagged as dropouts, and (6) measurements
  remaining after applying the mask.}
}
\end{deluxetable}

\begin{figure}
\figurenum{2}
\begin{center}
 \includegraphics[width=3.9in,angle=270]{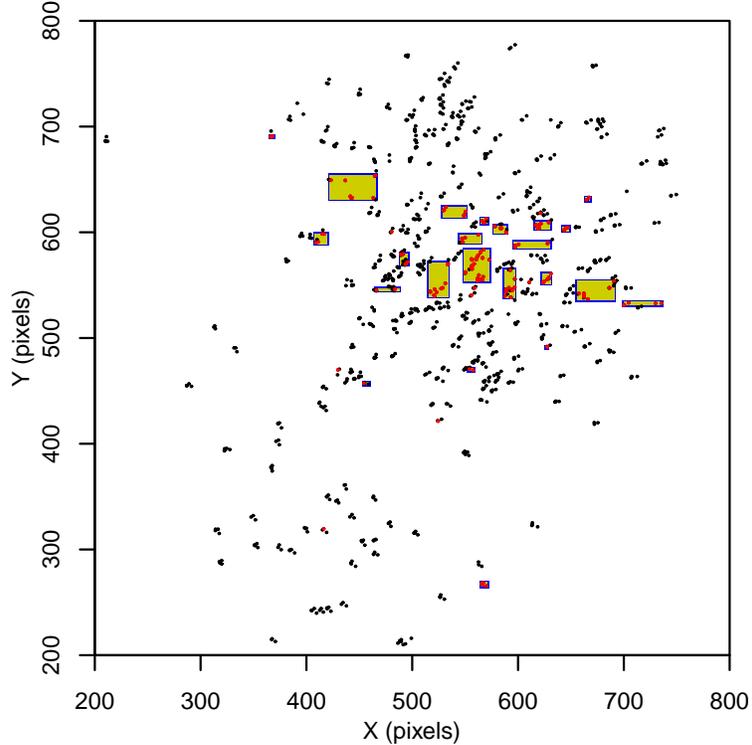} 
\caption{Mapping of dropout/non-dropout data onto the UVOT detector plane for the three UV bands.
Dropouts are plotted in red and non-dropouts in black.
Note that the dropouts cluster together in the detector plane.
The yellow rectangles show the 23 filtering boxes.
These boxes are outlined in blue to aid the eye, but the actual ``bad'' detector areas as shown by the regions shown in yellow.
Data from any of the four shortest-wavelength filters that fall within any of these boxes are excluded from the final light curves shown in Figure 3.}
\label{fig:fig2}
\end{center}
\end{figure}

As in the case for NGC~5548,
the dropouts are highly clustered in the detector plane.
\edit{We define 23 boxes to enclose clusters of three or more dropouts}
(Figure~\ref{fig:fig2}),
ranging in size from a single 1\arcsec$\times$1\arcsec\ pixel to over 1000 pixels.
We note that the data from both \ngc\ and NGC~5548 sample the detector plane sparsely and do not cover the exact same regions, so the mask defined for one of these AGN is not well-suited for the other.

\begin{figure}
\figurenum{3}
\begin{center}
\includegraphics[width=8.5in,angle=270]{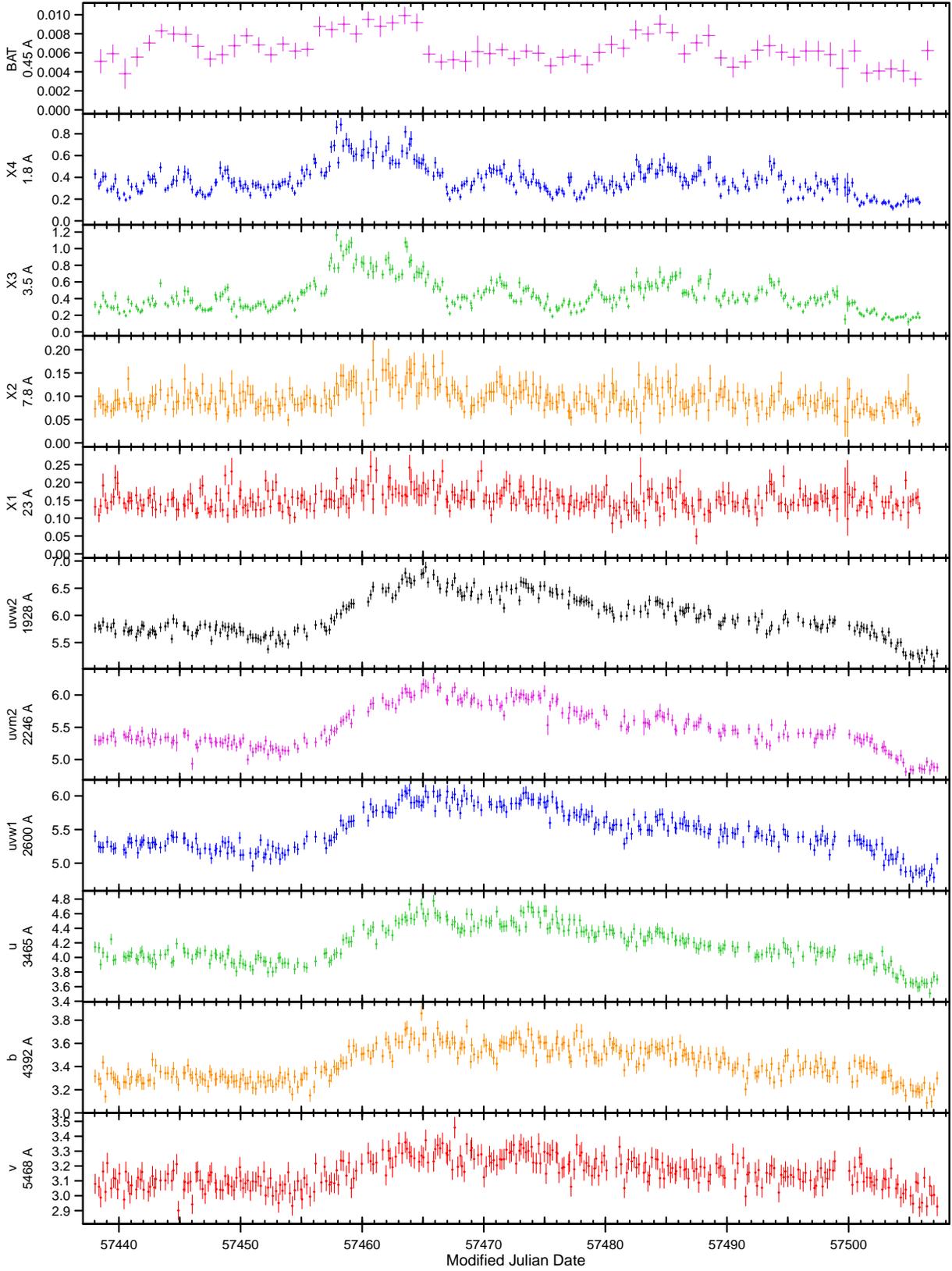} 
\caption{Light curves from the NGC 4151 \swift\ campaign.  
Data are ordered by wavelength, with the top panel from BAT (15--50 keV), the next four from XRT (X1\dots X4 = 0.3--1.25, 1.25--2.5, 2.5--5, 5--10 keV, respectively), and the bottom six from UVOT.
The plotted UVOT points are restricted to those that passed the filtering shown in Figures 1 and 2.
The X-ray data are all in units of ct s$^{-1}$ and the UVOT data are in units of $10^{-14}$ erg cm$^{-2}$ s$^{-1}$ \AA$^{-1}$.
The final seven XRT points were lost.}
\label{fig:fig3}
\end{center}
\end{figure}

The final step is to remove all data in the four shortest-wavelength bands that fell within any of these boxes.
\edit{The mask is not applied to} {\it b} or {\it v} data
because the effect is small compared to the statistical uncertainties in these bands.
\edit{Tallies of the points filtered out and of the final data points}
are given in Table~\ref{table2}.
Data that fall within the detector mask are shown as yellow Xs in Figure~\ref{fig:fig1}.
The final reduced and filtered UVOT data are given in Table~\ref{table3} and plotted in the lower six panels of Figure~\ref{fig:fig3}.

\begin{deluxetable}{lccc}
\tablecaption{UVOT Data \label{table3}}
\tablecolumns{4}
\tablehead{
\colhead{(1)} & \colhead{(2)} & \colhead{(3)} & \colhead{(4)} \cr
\colhead{MJD} & \colhead{Flux} & \colhead{Error}  & \colhead{Filter} \cr}
\startdata
57438.0447	&	5.767	&	0.080	&	{\it uvw2}	\cr
57438.3637	&	5.805	&	0.080	&	{\it uvw2}	\cr
57438.4962	&	5.755	&	0.080	&	{\it uvw2}	\cr
57438.6966	&	5.878	&	0.081	&	{\it uvw2}	\cr
57439.0400	&	5.788	&	0.080	&	{\it uvw2}	\cr
\enddata
\tablecomments{Column 1: modified Julian date at the midpoint of the exposure.
Column 2: measured flux in units of $10^{-14}$ erg cm$^{-2}$ s$^{-1}$ \AA$^{-1}$.
Column 3: measured 1$\sigma$ error in the same units.
Column 4: observing filter.
The data are sorted first by filter, then by MJD.
Only a portion of this table is shown here to demonstrate its form and content. 
A machine-readable version of the full table is available online.}
\end{deluxetable}

\subsection{XRT Data Reduction}
\label{section:xrt}

The \swift/XRT data were gathered in photon-counting (PC) mode for all except the last seven visits, which were gathered in WT mode owing to an error in our observing proposal.
The data were analyzed using the tools described by \cite{Evans09}\footnote{http://www.swift.ac.uk/user\_objects} to produce light curves that are fully corrected for instrumental effects such as pile up, dead regions on the CCD, and vignetting.
Because the WT data showed a large flux discontinuity with the PC data at soft energies, these seven WT points were discarded.
We additionally excluded all visits where the total good integration time was less than 120~s.
This resulted in a final light curve having 319 X-ray points over the 71 day monitoring period (see Table 1).

We generated X-ray light curves in four bands: X1 (0.3--1.25~keV), X2 (1.25--2.5~keV), X3 (2.5--5~keV) and X4 (5--10~keV).
These are chosen so each band spans one octave of frequency, except for X1, which was chosen to be larger because it also has the lowest count rate.
We utilize ``snapshot'' binning, which produces one bin for each spacecraft pointing. 
This is done because these short visits always occur completely within one orbit.
These XRT data are presented in Table~4.
As discussed in Section~2.1, the X1 band is only weakly variable because it is known to be dominated by extended emission \citep{Zdziarski02}.

\begin{center}
\begin{deluxetable*}{lcccccccc}
\label{table4}
\tablenum{4}
\tablecaption{XRT Data}
\tablecolumns{9}
\tablehead{
\colhead{(1)} & \colhead{(2)} & \colhead{(3)} & \colhead{(4)} & \colhead{(5)} & \colhead{(6)} & \colhead{(7)} & \colhead{(8)} &  \colhead{(9)} \cr
\colhead{MJD} & \colhead{X1 Flux} & \colhead{X1 Error} 
 & \colhead{X2 Flux} & \colhead{X2 Error} 
 & \colhead{X3 Flux} & \colhead{X3 Error} 
 & \colhead{X4 Flux} & \colhead{X4 Error} \cr}
\startdata
57438.0435	&	0.132	&	0.024	&	0.073	&	0.018	&	0.328	&	0.038	&	0.430	&	0.044	\cr
57438.3631	&	0.108	&	0.019	&	0.100	&	0.018	&	0.236	&	0.028	&	0.322	&	0.033	\cr
57438.4963	&	0.134	&	0.020	&	0.090	&	0.016	&	0.304	&	0.030	&	0.358	&	0.033	\cr
57438.6955	&	0.192	&	0.032	&	0.082	&	0.021	&	0.436	&	0.048	&	0.405	&	0.046	\cr
57438.8985	&	0.150	&	0.021	&	0.071	&	0.014	&	0.361	&	0.032	&	0.408	&	0.034	\cr
\enddata
\tablecomments{Column 1: modified Julian date.
Columns 2 and 3: measured X1 flux and 1$\sigma$ error, in ct sec$^{-1}$.
Columns 4 and 5: measured X2 flux and 1$\sigma$ error, in ct sec$^{-1}$.
Columns 6 and 7: measured X3 flux and 1$\sigma$ error, in ct sec$^{-1}$.
Columns 8 and 9: measured X4 flux and 1$\sigma$ error, in ct sec$^{-1}$.
Only a portion of this table is shown here to demonstrate its form and content. 
A machine-readable version of the full table is available online.}
\end{deluxetable*}
\end{center}

\subsection{BAT Data Reduction}
\label{section:bat}

Besides the pointed UVOT and XRT instruments, \swift\ also has the BAT, a large-sky monitor originally developed to pinpoint new $\gamma$-ray bursts \citep{Barthelmy05}.
The BAT is now also being used to monitor X-ray transients in the hard X-ray (15--50~keV) band (e.g., \citealt{Krimm13}).
Most AGN are too faint to produce usable high-cadence BAT light curves.
However, \ngc\ is typically the brightest Seyfert~1 in the sky at these wavelengths, so we were able to utilize these data\footnote{Current BAT data for \ngc\ are available at http://swift.gsfc.nasa.gov/results/transients/weak/NGC4151.lc.txt} to measure a hard X-ray light curve.
This provides an important extension of the XRT light curves to higher energies.
These data are reproduced in Table~5.
See \cite{Krimm13} for further details of the BAT data gathering and reduction process.

\begin{deluxetable}{lcc}
\label{table5}
\tablenum{5}
\tablecaption{BAT Data}
\tablecolumns{5}
\tablehead{
\colhead{(1)} & \colhead{(2)} & \colhead{(3)} \cr
\colhead{MJD} & \colhead{BAT Flux} & \colhead{BAT Error} \cr}
\startdata
57438.5	&	0.00510	&	0.00124	\cr
57439.5	&	0.00591	&	0.00094	\cr
57440.5	&	0.00380	&	0.00157	\cr
57441.5	&	0.00555	&	0.00099	\cr
57442.5	&	0.00703	&	0.00080	\cr
\enddata
\tablecomments{Column 1: modified Julian date.
Columns 2 and 3: measured BAT flux and 1$\sigma$ error, in ct s$^{-1}$.
Only a portion of this table is shown here to demonstrate its form and content. 
A machine-readable version of the full table is available online.}
\end{deluxetable}

\subsection{Light Curves}
\label{section:lcs}

Figure~3 shows the resulting light curves, presented in order of descending frequency with the highest frequency band at the top and the lowest at the bottom.
These data are unprecedented in two respects.
First, the average sampling interval of 0.22--0.28 day is a factor of $\sim$2 better than that obtained for the \swift\ monitoring of NGC~5548, the previous most-intensive AGN monitoring of this type \citep{Edelson15}.
Second, because \ngc\ is typically the brightest Seyfert~1 in the sky, it was possible to measure five independent X-ray bands, including the hard X-rays with BAT.
All 11 of the resulting light curves in Figure~3 are used for time-series analysis.

\edit{The visual impression of the UV/optical ({\it uvw2} through {\it v}) light curves is that they are so similar that the variations clearly appear to be related.
The same is true for the relatively hard X-rays (BAT through X3).
The X2 light curve has a lower S/N and may be related to the harder X-rays while the X1 light curve shows almost no detectable signal so its relation to other bands cannot be assessed.
Comparison of the BAT--X3 bands with the {\it uvw2--v} bands shows that while many of the largest peaks seen in the X-rays also appear in the UV/optical, the character of the variations is not identical, in the sense that the most rapid variations seen in the X-rays are not seen in the UV/optical.  
This could be due to smoothing and lagging of the X-ray light curves to produce the UV/optical light curves or it could be that the two wavelength regimes simply have different drivers and the apparent long-term similarities are just a chance coincidence due to the red-noise character of AGN variability \citep{Vaughan03}. 
This caveat that the X-rays may not be driving the UV/optical should be kept in mind throughout this analysis.
This possibility will be assessed further in subsequent papers (e.g. K. Horne et al. in preparation).}

\section{Time-series Analysis}
\label{section:tsa}

\subsection{Cross-correlation-Functions}
\label{section:ccf}

The focus of this paper is on testing and constraining continuum-emission models primarily through measurement of interband lags.
We used the interpolated cross-correlation function (CCF) as implemented by \cite{Peterson04} to measure and characterize the temporal correlations and interband lags within these data.\footnote{	The code, called {\tt sour}, used to compute CCFs is available at \url{https://github.com/svdataman/sour}}

We first normalized the data by subtracting the mean and dividing by the standard deviation.
These were derived ``locally'' --- only the portions of the light curves that are overlapping for a given lag are used to compute these quantities. 
We implemented ``2-way'' interpolation, which means that for each pair of bands we first interpolated in the ``reference'' band and then measured the correlation function, next interpolated in the ``subsidiary'' band and measured the correlation, and subsequently averaged the two to produce the final CCF.

We then used the ``Flux Randomization/Random Subset Selection'' (FR/RSS) method \citep{Peterson98} to estimate uncertainties on the measured lags.
This is a Monte Carlo technique in which lags are measured from multiple realizations of the CCF.
The FR aspect of this technique perturbs \edit{in a given realization} each flux point consistent with the quoted errors \edit{assuming a Gaussian distribution of errors.
In addition,} for a time series with $N$ data points, the RSS randomly draws with replacement $N$ points from the time series to create a new time series.
In that new time series, the data points selected more than once have their error bars decreased by a factor of $n_{\rm rep}^{-1/2}$, where  $n_{\rm rep}$ is the number of repeated points. 
Typically a fraction of $1/e$ of data points are not selected for each RSS realization. 
In this paper, the FR/RSS is applied to both \edit{the ''driving'' and ''responding''}
light curves in each CCF pair.
The CCF (r($\tau$)) is then measured and a lag determined to be the weighted mean of all points with $ r > 0.8 r_\mathrm{max} $, where $ r_\mathrm{max} $ is the maximum value obtained for $r$, given in Column 2 of Tables 5 and 6.
For the data presented herein, simulated lags are determined for \edit{200,000} realizations and then used to derive 68\% confidence intervals.

Because of the wealth of \swift\ data, covering 11 bands, we perform two complete CCF analyses: one with the UVOT band {\it uvw2} as a reference \edit{(Table~6)} and the other with the XRT band X3 as the reference \edit{(Table~7)}.
This allows for a more sensitive search for small lags within the UV/optical and X-rays as well as between the UV/optical and X-ray regimes than would be possible with just a single set of CCFs.
These results are presented graphically in Figure~4 and listed in Tables~6 and 7.
We first describe the results relative to the X3 X-ray band and then to the {\it uvw2} UV band.

\begin{deluxetable}{lccccc}
\label{table6}
\tablenum{6}
\tablecaption{{\it uvw2} CCF Results}
\tablewidth{0pt}
\tablecolumns{6}
\tablehead{
\colhead{(1)} & \colhead{(2)} & \colhead{(3)} & \colhead{(4)} 
	& \colhead{(5)} & \colhead{(6)} \cr
\colhead{Band} & \colhead{$r_{\rm max}$} & \colhead{$\tau_{\rm med}$ (days)} 
 & \colhead{$\tau_l$ (days)} & \colhead{$\tau_u$ (days)} 
 & \colhead{Sig. (\%)} \cr}
\startdata
BAT		&	0.46	&	-7.42	&	-13.49 &	-3.08	   & 77.2 \cr
X4		&	0.64	&	-3.58	&	-4.04	&	-3.22	     & 88.6 \cr
X3		&	0.68	&	-3.39	&	-3.72	&	-3.11	     & 90.8 \cr
X2		&	0.56	&	-2.28	&	-3.10	&	-1.58	     & 83.1 \cr
X1		&	0.33	&	-3.74	&	-8.26	&	-1.55	     & 68.8 \cr
{\it	uvw2}	&	1.00	&	0.00	&	-0.25	&	0.24 & $>$99.9 \cr
{\it	uvm2}	&	0.97	&	0.01	&	-0.21	&	0.26 & $>$99.9 \cr
{\it	uvw1}	&	0.95	&	0.02	&	-0.24	&	0.29 & $>$99.9 \cr
{\it	u}	&	0.95	&	0.61	&	0.33	&	0.88	 & $>$99.9 \cr
{\it	b}	&	0.89	&	0.83	&	0.49	&	1.15	 & $>$99.9 \cr
{\it	v}	&	0.82	&	0.96	&	0.50	&	1.43	 & $>$99.9 \cr
\enddata
\tablecomments{Column 1: band.
Column 2: maximum correlation coefficient.
Column 3: median lag.
Columns 4 and 5: 68\% confidence interval lower and upper limits.
\edit{Column 6: peak significance estimate.}
Note that all correlations are measured relative to band {\it uvw2}, so the sixth line refers to the autocorrelation, all others are cross-correlations.
}
\end{deluxetable}

\begin{deluxetable}{lccccc}
\label{table7}
\tablenum{7}
\tablecaption{X3 CCF Results}
\tablewidth{0pt}
\tablecolumns{6}
\tablehead{
\colhead{(1)} & \colhead{(2)} & \colhead{(3)} & \colhead{(4)} 
	& \colhead{(5)} & \colhead{(6)} \cr
\colhead{Band} & \colhead{$r_{\rm max}$} & \colhead{$\tau_{\rm med}$ (days)} 
 & \colhead{$\tau_l$ (days)} & \colhead{$\tau_u$ (days)}
 & \colhead{Sig. (\%)} \cr}
\startdata
BAT		&	0.75	&	-0.42	&	-1.65	&	4.19       & 99.9 \cr
X4		&	0.92	&	-0.10	&	-0.24	&	0.04	     & $>$99.9 \cr
X3		&	1.00	&	0.00	&	-0.14	&	0.14	     & $>$99.9 \cr
X2		&	0.57	&	1.35	&	0.89	&	1.78	     & 99.2 \cr
X1		&	0.34	&	1.07	&	-8.31	&	3.35	     & 93.8 \cr
{\it	uvw2}	&	0.68	&	3.40	&	3.12	&	3.72 & 91.4 \cr
{\it	uvm2}	&	0.67	&	3.64	&	3.30	&	4.07 & 89.8 \cr
{\it	uvw1}	&	0.64	&	3.68	&	3.23	&	4.09 & 87.4 \cr
{\it	u}	&	0.60	&	3.63	&	3.22	&	4.17 	 & 85.1 \cr
{\it	b}	&	0.58	&	3.13	&	2.67	&	3.69	 & 88.5 \cr
{\it	v}	&	0.56	&	4.16	&	3.28	&	5.25	 & 91.0 \cr
\enddata
\tablecomments{Column 1: band.
Column 2: maximum correlation coefficient.
Column 3: median lag.
Columns 4 and 5: 68\% confidence interval lower and upper limits.
\edit{Column 6: peak significance estimate.}
Note that all correlations are measured relative to band X3, so the third line refers to the autocorrelation, and all others are cross-correlations.
}
\end{deluxetable}

Because \ngc\ is so bright, we are able to measure the BAT/X3 CCF.
It shows a correlation consistent with zero lag, though the confidence interval is \edit{much larger than with the other X-ray bands,} owing to the relatively poor sampling and S/N of BAT compared to the XRT bands.
Likewise, the X4/X3 correlation is very strong and consistent with zero lag, with much tighter limits on the lag.
The X2/X3 correlation is weaker but still apparently significant, with an $\sim$1.5 day lag in the sense that the harder band leads the softer band.
Because the X1 variations are very weak, consistent with an origin in an extended region (e.g., \citealt{Zdziarski02}; \citealt{Keck15}), the X1/X3 CCF indicates at best a weak correlation, making the delay measurement much less accurate than for the other bands or perhaps not measurable at all.
That correlation does not appear significant and no meaningful lag could be measured with that band.
All six UVOT bands are strongly correlated with X3, with well-detected lags in the sense that the X-rays lead the UV/optical by $\sim$3--4 days.

The {\it uvw2}-referenced results are more sensitive to lags within the UV/optical than are the X3-referenced results.
These show very strong correlations within the UV with no measurable lags down to limits of $\sim$0.5~day, while the optical bands appear to show small but possibly significant ($\sim 2\sigma$) lags of $\sim$0.6--1~day behind {\it uvw2}.
As with the X3-referenced CCFs, these data also show no correlation with the softest X-ray band (X1) and significant correlations with bands X2--X4, with the lag to the X2 band ($\sim$2 days) midway to the lags with X3 and X4 (3--4 days).
The BAT/{\it uvw2} correlation is not significant, probably owing in part to the relatively poor sampling and S/N of the BAT light curve.

\subsection{\edit{CCF Significance Testing}}
\label{section:sig}

\edit{We now evaluate the significance of the observed CCF peaks by estimating the probability that the observed $r_\mathrm{max}$ could arise by chance from independent AGN-like light curves.
This is based on the Monte Carlo simulation technique of (\citealt{Breedt09}; see also \citealt{Breedt10}, \citealt{Cameron12}), which cross-correlates the higher-energy observed light curve against simulated observations for a large number of independent AGN-like light curves, each of which are constructed to match the mean and rms of the log(flux) of the lower-energy light curve.
Initially, the slope of the power spectral density (PSD) function of the ''driving'' band (always assumed to be the highest energy of the two) is estimated from the data with a single power-law fit to that PSD. 
The probability density function of the flux of the driving band is also fitted using a lognormal distribution and used to simulate light curves by the method of \cite{Emmanoupolous13}, as implemented by \cite{Connolly15}, with an appropriate level of measurement noise added back in.
Then the simulated light curve is cross-correlated with the actual lower-energy light curve and tested to see if the highest observed value of the correlation coefficient ($r_\mathrm{max}$) exceeded that of the actual data.
This process is repeated 10,000 times for each CCF band pair, with a new simulated light curve leading to a new CCF in each realization.
This  ensemble of simulated CCFs should by construction contain no real correlated signal, allowing us to calculate the probability of finding a particular correlation coefficient at a given lag time by chance.}

\edit{Our initial testing indicated that for low S/N light curves (e.g. X1) the method used a flat (nearly white [measurement] noise) PSD to generate the synthetic light curves, resulting in implausibly high significances.
For instance initial application of this technique yielded 99.6\% significance for the X1/{\it uvw2} CCF peak (which had $r_\mathrm{max} \sim 0.33$) and 87\% for the X3/{\it uvw2} peak ($r_\mathrm{max} \sim 0.68$).
In many cases (e.g. X1 and BAT), the data are inadequate to measure even the a basic PSD slope.
We therefore investigated adapting this procedure to use synthetic PSD slopes plus noise instead of attempting to measure them from the data.}

\edit{We ran simulations assuming PSD slopes of -3, -2.5 and -2 to cover most of the observed range of AGN behavior (e.g. \citealt{Gonzalez12}, 
\citealt{Edelson14}).}
\edt{Only pure power-law PSD slopes were assumed in this initial analysis; no broken or bending power laws were used.}
\edit{Furthermore, this new approach compares the observed value of $r_\mathrm{max}$ to that derived from the simulations across a window of $\pm$10 days, whereas we previously only compared the observed value of $r_\mathrm{max}$ with the simulated value of $r$ at that same lag.
(This is done because we do not have an a priori idea of where the peak will fall.)
In this case, because we tested three PSDs with 30,000 runs each, a total of 90,000 runs were used for each CCF band pair.}
\edt{For consistency, we use this revised technique on all CCFs in this paper.}
\edit{These results are shown in Column~6 of Tables ~6 and 7.}
\edt{No more than marginal differences were found for simulations with different PSD slopes.}
\edit{Note, in particular, that the X1/{\it uvw2} CCF significance is now much lower ($\sim$69\%, consistent with no correlation), }
\edt{but the X3/{\it uvw2} significance has risen to $\sim$91\%.
This is all}
\edit{as expected.
Still, these results should be considered preliminary because thorough analysis of this significance test (e.g. testing a wider range of PSD slopes, testing bending PSDs, varying the window size) has not been completed.
That is beyond the scope of this paper, but will be presented in a future paper (S. Connolly et al. in prep).}

\subsection{Interband Lag Fits}
\label{section:fits}

In this section, we use these CCF results to establish the relation between lag and wavelength, using a methodology similar to that of \cite{Edelson15}.
The {\it uvw2}-referenced CCFs are used because those are the most sensitive to small lags within the UV/optical.
\edit{The BAT and X1 lags are excluded from the analysis because of the relatively low significance of the correlation peak ($< 1 \sigma $) and large errors on the lag confidence intervals ($\sim 6-10$~days).}
The {\it u}-band lag is also ignored because of possible contamination of diffuse continuum emission from BLR clouds, which must be present at some level and is expected to be stronger in this band \citep{Korista01}.
The remaining three X-ray and five UVOT lags are shown as a function of the observing band central wavelength in Figure~5.
These data were then modeled with a function of the form 
\edit{\begin{equation}\label{eq:}
\tau = \tau_0 [ (\lambda/\lambda_0)^{4/3} - 1 ], 
\end{equation}}
where $\lambda_0= 1928 $~\AA, the wavelength of the reference {\it uvw2} band
\edit{and $\tau_0$ is effectively the fitted lag between wavelength zero and $\lambda_0$ in days.}
The {\it uvw2} autocorrelation function lag is identically zero, so this point does not participate in the fit but instead the fit is forced to pass through this point.

If the X-rays originate near the black hole and drive emission at longer wavelengths, then the X-ray lag represents the light-travel time from the center of the system. 
In this way, the X-ray data anchor the zero point of the fit and set the physical size scale of the disk.
This is the simplest functional form consistent with the standard reprocessing model, yet the data are not well-fitted by this function, with reduced $\chi^2$ of 139.6 for 6 degrees of freedom (dof) for an unacceptable $p$-value $<10^{-27}$.
In particular, this shows that it is impossible to  simultaneously fit the X-ray and UV/optical data points due to the $\sim$3 day lag between variations in these regimes.

The results are quite different if the three X-ray points are excluded from the analysis.
Modeling the same function to just the five UVOT points yields a reduced $\chi^2$ of 0.59/3 dof, for an acceptable $p$-value of 0.90.
If this fit is extrapolated to zero at the inner edge of the disk, then the fit parameter $ \tau_0 = 0.34 \pm 0.11 $~day would indicate that emission from an annulus $\sim$0.34 lt-day in radius peaks at 1928~\AA.
Note, however, that the X-ray points are inconsistent with this extrapolation, because the X4, X3, and X2 lags undershoot the fit by 7.7$\sigma$, 9.9$\sigma$, and 2.5$\sigma$, respectively.
This indicates that the assumption underlying the reprocessing model, that all interband lags are caused by light-travel time effects, cannot be correct.

\section{Discussion}
\label{section:rep}

\subsection{\edit{Variability Timescales}}
\label{section:ts}

\edit{This analysis of the \ngc\ \swift\ data indicates (1) a clear $\sim$3 day lag between X-ray and UV/optical variations, (2) smaller $\sim$0.5-1 day lags within the UV/optical, and (3) at lower confidence, a possible $\sim$1.5 day lag between relatively hard X-ray (2.5-10~keV) and softer X-ray (1.25-2.5~keV) band variations.
Before using these results to test models, we examine size scales that would be related to these timescales under a variety of general assumptions.}

\edit{The most basic size scale is the light-crossing size.  
We assume that \ngc\ has a black hole mass of $ 4 \times 10^7 $~M$_\odot$ (\citealt{Bentz06}; \citealt{Onken14}), so the gravitational radius $r_g = 200 $~lt-s = 0.0023 lt-day.
Thus a lag of 1.5-3~lt-day corresponds to a light-crossing size of $ R \sim $650-1300~$r_g$.
This is significantly larger than the expected size of the inner accretion disk/corona region, so we conclude that the observed lags do not correspond to light travel sizes within the central engine.}

\edit{The dynamical timescale, over which a vertical disturbance in a disk returns to hydrostatic equilibrium, is given by $ t_\mathrm{dyn} \sim (R^3/GM)^{1/2}$ \citep{King08}.
Again for a mass of $ 4 \times 10^7 $~M$_\odot$, a delay of 1.5-3 days corresponds to a region of size $ R = 23-36 r_g $ for re-establishing hydrostatic equilibrium.
Furthermore, the thermal ($t_\mathrm{th}$) and viscous ($t_\mathrm{visc}$) timescales are related to the dynamical timescale by 
$ t_\mathrm{dyn} \sim \alpha t_\mathrm{th} \sim \alpha (H/R)^2 t_\mathrm{visc}$,
where $\alpha$ is the dimensionless viscosity parameter \citep{Shakura73}, thought to be of the order of 0.1 in AGN, and $H/R$ is the ratio of height to radius of the disk, also of the order of 0.1.
This means that for both thermal and viscous processes, a delay of 1.5-3 days would correspond to a region smaller than the last stable orbit around the black hole (6$r_g$).}

\edit{Thus we conclude that the observed lags of 1.5-3 days cannot be associated with any plausible light-crossing time (that is too large) or thermal or viscous processes (those timescales are too small). 
However, they may be associated with a process governed by the dynamical timescale ($t_\mathrm{dyn}$), which is also the timescale on which the disk responds to loss of hydrostatic equilibrium \citep{Gardner17}.
A possible implication of associating the observed lags with the dynamical timescale is discussed in Section~4.4.}

\subsection{Accretion Disk Size}
\label{section:disk}

In this subsection, we compare the size of the accretion disk in \ngc\ derived from RM and the reprocessing model with theoretical predictions.
We start with Equation 12 of \cite{Fausnaugh16}, which gives the light-crossing radius $r$ of a \edit{flat}, geometrically thin, optically thick accretion disk annulus emitting at a characteristic wavelength $\lambda_0$:
\begin{equation}\label{eq:}
r = \left(X {k \lambda_0 \over hc}\right)^{4/3}
\left[\left({GM \over 8 \pi \sigma} \right)
\left({L_\mathrm{Edd} \over \eta c^2} \right)
(3 + \kappa) \dot{m}_\mathrm{E} \right]^{1/3}
\end{equation}
where $X$ is a multiplicative scaling factor of order unity that accounts for systematic issues in converting the annulus temperature $T$ to wavelength $\lambda$ at a characteristic radius $R$,
$L_\mathrm{Edd}$ is the Eddington luminosity, 
$\eta$ is the radiative efficiency in converting mass into energy,
$\kappa$ is the local ratio of external to internal heating, assumed to be constant with radius,
and the Eddington ratio $ \dot{m}_\mathrm{Edd} = L_\mathrm{bol} / L_\mathrm{Edd} $.
Under the assumption that at an annulus of radius $R$ the observed wavelength corresponds to the temperature given by Wien's Law, then $ X = 4.87 $ 
If instead the flux-weighted radius $\langle R \rangle$ is used, then $ X = 2.49 $.
($ \langle R \rangle = \int_{R_0}^{\infty} B(T(R))R^2 \, dR / \int_{R_0}^{\infty} B(T(R)) R \, dR $, where $R_0$ is the inner edge of the disk, $B(T(R))$ is the Planck function, and $T(R)$ is the temperature at radius $R$.)
The flux-weighted estimate assumes that the temperature profile of the disk is described by $ T \propto R^{-3/4} $ \citep{Shakura73}.
In both the Wien and flux-weighted cases, the disk is assumed to have a fixed aspect ratio and to be heated internally by viscous dissipation and externally by the coronal X-ray source extending above the disk (the lamp-post model).

Assuming $ \kappa = 0 $ (negligible external heating compared to internal heating) and $ \eta = 0.1 $ and setting the fiducial wavelength $\lambda_0 = 1928 $~\AA\ yields a more convenient scaling:
\begin{equation}\label{eq:}
r = ct = 0.09 \left(X\frac{\lambda}{1928{\rm \AA} } \right)^{4/3}\nonumber M_8^{2/3} \left(\frac{\dot m_{\rm Edd}}{0.10} \right)^{1/3}
\textrm{lt-dy}
\end{equation}
where $r$ is given in units of light days and $M_8 = M_{\rm BH} / 10^{8}{\rm M_{\odot}} $.

We use as the black hole mass of \ngc\ $ M_8 = 0.4 $ as discussed above.
The Eddington ratio is difficult to estimate, and \ngc\ has a wide variety of estimates including 6\% \citep{Meyer11}, 4\% \citep{Kraemer05}, and 0.6\% \citep{Edelson96}.
The first two values were estimated assuming $ M_8 = 0.13 $ while the last assumed $ M_8 = 0.4 $.
Correcting the first two values to a mass of $ M_8 = 0.4 $ as used herein yields $\dot m_\mathrm{Edd} = 2$\%, 1.3\%, and 0.6\%, respectively.
Here we assume a value of 1\%, close to the harmonic mean of these estimates.

Inputting these values into Equation 3 yields $ r = 0.19$ light-day for the Wein's Law case and $ r = 0.08$ light-day for the flux-weighted assumption.
Given the fitted value of $ t = 0.34 \pm 0.11 $~day, the observed size appears larger than predicted by a factor of $\sim 2$-4.
A similar discrepancy was found between the RM-derived and theoretically predicted size of the accretion disk of the Seyfert 1 galaxy NGC~5548 (\citealt{Edelson15}, \citealt{Fausnaugh16}).
Gravitational microlensing of much more distant quasars also appears to derive disk sizes that are larger than predicted (e.g., \citealt{Morgan10}).

However, we must note that this particular disk size discrepancy should not be seen as highly significant for two reasons.
First, there are large systematic uncertainties in many of the input parameters in Equation 2.  
As mentioned earlier the historically derived black hole mass of \ngc\ ranges over a factor of $\sim$3, and even after correcting the Eddington ratio estimates to the same mass, those also range over an additional factor of $\sim$3.
In addition the radiative efficiency $\eta$ and ratio of external to internal heating $\kappa$ are also not well established in AGN in general or \ngc\ in particular.
Thus the theoretically expected value of $r$ is not very well determined for this object.

Second, the \swift\ data alone do not provide strong constraints on the observed light-crossing time $t$ because of the limited wavelength range and the poor S/N, and the weak variability in {\it v} band in particular.
A much stronger test of the UV/optical data's consistency with the thin-disk model will be possible using the simultaneous ground-based data, which go all the way to the {\it z} band ($\sim$9000~\AA), at much higher S/N.
Those data will be presented and this test will be performed in a future paper (K. Horne et al., in preparation).

\subsection{Implications for the Lamp-Post Model}
\label{section:imp}

The observed $ \tau \propto \lambda^{4/3} $ relation and $\sim$0.5--1 day lags within the UV/optical are consistent with the standard thin-disk picture \citep{Shakura73}, albeit with large uncertainties.
However, the full lag-wavelength relation shown in Figure~5, including the X-ray lags, indicates that these results are in fundamental disagreement with the standard lamp-post reprocessing model in which the small X-ray emitting corona directly illuminates and drives variations in the extended UV/optical-emitting accretion disk.
The $\sim$3 day lag between the hard X-rays and UV and the $\sim$0.5--1 day lags within the UV/optical cannot be simultaneously explained in terms of direct illumination of the disk by the corona.

\edit{The UV/optical light curves also appear to be considerably smoother on short timescales than the X-ray light curves. 
This problem was noted by \cite{Gardner17} with regards to the NGC 5548 data, where the UV/optical lightcurves similarly lack the high frequency power seen in the X-rays. 
However, the NGC~5548 X-ray S/N was much worse than for \ngc\ (which in most wavelength bands is the brightest Type 1 AGN in the sky), which may be why the NGC~5548 data do not require a long X-ray/UV lag.}
In retrospect, this effect \edit{can be seen, to some degree,} in earlier \swift\ monitoring, such as of MR2251-178 \citep{Arevalo08} and NGC~2617 \citep{Shappee14}.
In all of these cases, it appears that there is no simple way to reconcile the relatively small lags of the UV/optical with the large X-ray/UV lags within the context of the direct reprocessing lamp-post model.

\subsection{Possible Alternatives}
\label{section:alt}

\cite{Gardner17} developed a picture to explain the poor correlation between the X-ray and UV/optical variability in NGC 5548. 
That same picture also provides a natural explanation for the wavelength-lag structures seen in \ngc, in particular the observed relatively long 3-4~day X-ray/UV lag and the smaller $\lesssim$1~day lags within the UV/optical. 
This is done by invoking an additional component that emits in the EUV and acts as an intermediary between the corona and disk.
This putative EUV component offers an explanation for the origin of both the ``big blue bump'' and ``soft X-ray excess'' as low- and high-energy tails of a component that peaks in the intrinsically unobservable EUV spectral region.

The key to this picture is that instead of the X-ray corona directly illuminating the disk that then processes and re-emits the energy in the UV/optical, two separate reprocessings occur: first, the corona illuminates and heats the EUV component, which is smaller than the disk (thus much smaller than light days in size), so the first reprocessing must occur on a timescale longer than the light-crossing size of the EUV component.
A sketch of how these components may be arranged physically is shown in Figure~6.
Hence, the X-ray/UV lag indicates some slower physical process.  
\edit{This would introduce both a lag and smoothing between the X-ray and UV/optical bands, as has been observed.
As discussed earlier, a lag time of $\sim$1.5-3 days would indicate a size of $23-36 r_g$ for the inner and outer radii of the EUV torus if it was associated with the dynamical timescale.
In this model the heating would cause the torus to puff up in the vertical direction, so a dynamical timescale (the time required for the system to return to relax back in the vertical direction) would be naturally associated with this process.}

\edit{Then, in the standard thin-disk picture, a second reprocessing would occur when the EUV torus illuminates and heats the accretion disk on the light-crossing time, which then radiates the observed UV/optical radiation.
An alternative scenario, envisioned in the \cite{Gardner17} picture,  is that the inner edge of the disk responds to an increase in heating, caused by increased illumination from the outer edge of the EUV torus, by expanding upward on the dynamical timescale.
As a result, the inner disk radii are continually transitioning between a standard thin-disk state and a larger scale height state, which is more similar to that of the material in the EUV torus. 
This inward/outward pulsating of the EUV torus-standard disk boundary, in response to the X-ray heating of EUV torus inner edge, is potentially a cause of the UV/optical lags in the \cite{Gardner17} picture.}

We emphasize that at this early stage other alternatives are certainly possible.
For instance, \cite{Korista01} find that the diffuse continuum from the BLR is expected to contribute to the measured UV/optical lag-wavelength relation, even broadly mimicking the increasing lag with wavelength behavior. 
However, the diffuse continuum from the BLR is unlikely to be the sole source of the observed continuum lag spectrum, and it probably cannot explain the mismatch in lags between the UV/optical and the X-ray continuum bands observed in this object and others.

\section{Conclusions}
\label{section:concl}

A 69 day \swift\ monitoring campaign yielded light curves in six UV/optical bands covering 1900--5500~\AA\ and five X-ray bands spanning 0.3--50 keV.
CCF analysis shows the UV/optical variations are strongly correlated with a small $\sim$1~day lag between the shortest and longest wavelengths.
The hard X-rays ($\sim$2.5--50~keV) are also strongly correlated, but there is a clear UV/optical lag of $\sim$3 days relative to the X-rays.
This does not appear consistent with the standard reprocessing models in which an X-ray emitting corona directly illuminates and drives variations in a standard \citep{Shakura73} thin accretion disk.
Instead, these results are broadly consistent with the existence of a second reprocessor that emits in the EUV \citep{Gardner17}.
The process by which the corona heats the putative EUV component appears to be much slower than simple light-travel time.
This EUV torus then apparently illuminates and heats the disk on the light-crossing timescale as in the standard reprocessing model.
Modeling of these data based on the work of \cite{Gardner17} will be addressed in a future work in which the goal will be to confirm or refute the hypothesis that such a ``double reprocessing'' model can explain these observations. 

While this experiment has yielded the clearest evidence to date for the EUV emission component, it does not strongly constrain the disk parameters because the longest-wavelength band sampled by \swift\ ({\it v}) has a very poor S/N and weak variability.
However, an intensive ground-based campaign has gathered simultaneous ground-based photometry on \ngc\ out to $\sim$9000~\AA\ ({\it z} band).
These data will be presented in a future paper (K. Horne \et, in preparation), as will a similar simultaneous spectroscopic monitoring campaign (M. Goad \et, in preparation).
These data should yield a much clearer picture of the structure and physical conditions of the disk and the larger BLR.

This \edit{experiment} and the previous \swift\ monitoring of NGC~5548 have opened up a new technique for studying the central regions of AGN in which the bulk of the luminosity is produced and emitted.
A third campaign on the low-luminosity Seyfert 1 galaxy NGC~4593 has just been completed, and those data are now being analyzed and prepared for publication (I. McHardy \et, in preparation).
In all three of these sources, the X-rays contribute a larger fraction of bolometric luminosity than is typical for AGN with higher luminosity and Eddington ratio AGN.  
It is certainly possible that AGN with more typical properties will behave differently.  
Thus we note that in the coming year disk RM will be performed on Mrk~509 and Mrk~110, which have much higher luminosity and Eddington ratio, respectively, than any of these AGN.
It is worth noting that \swift, a satellite originally designed to study $\gamma$-ray bursts, is now providing insights that no other observatory could into the structure and physical conditions in the central engines of AGN.


\acknowledgments 
\edit{We dedicate this paper to the memory of Neil Gehrels, the P.I. of \swift\ and a leading author of this paper.
Neil was a great scientist who also brought out the best in others.
He led the \swift\ team with enthusiasm and expertise, always happy and eager to take the satellite in new directions. 
For example, his strong and unwavering support is what allowed \swift\ to gather these unprecedented data on NGC~5548 and \ngc, providing a powerful method that should continue to inform our understanding of AGN physics for years to come.
Specifically, without his courageous approval of increased UVOT filter changes well beyond the design lifetime of the filter wheel, the extraordinary 11-band light curve in Figure 3, which forms the basis of this work, could not have been gathered.
This is but one tiny piece of the rich legacy that he leaves us.}

We \edit{further} thank the \edit{entire} \swift\ team for their tireless dedication and flawless execution of these programs\edit{ and acknowledge} the use of public data from the \swift\ data archive.
We also thank the anonymous referee for timely and helpful reports.
This research has made use of the XRT Data Analysis Software (XRTDAS) developed under the responsibility of the ASI Science Data Center (ASDC), Italy.
R.E. gratefully acknowledges support from NASA under awards NNX13AC26G, NNX13AC63G, and NNX13AE99G.
J.M.G. acknowledges support from NASA under award NNH13CH61C.
Research by A.J.B. and H.A.V. was supported by NSF grant AST-1412693.
\edit{K.H. acknowledges support from STFC grant ST/M001296/1.}
\edit{M.B. gratefully acknowledges support from the NSF through CAREER grant AST-1253702.}
\edit{S.M.C.} and E.R.C. gratefully acknowledge the receipt of research grants from the National Research Foundation (NRF) of South Africa.
P.E. and K.P. acknowledge support from the UK Space Agency.
A.V.F. and W.Z. are grateful for financial assistance from NSF grant AST-1211916, the TABASGO Foundation, and the Christopher R. Redlich Fund.
D.C.L. gratefully acknowledges support from NSF under grants AST-1009571 and AST-1210311.
T.T. has been supported by NSF grant AST-1412315 and by the Packard Foundation in the form of a Packard Research Fellowship.
M.V. gratefully acknowledges support from the Danish Council for Independent Research via grant no. DFF 4002-00275.


\clearpage

\begin{figure}
\figurenum{4}
\begin{center}
\includegraphics[width=8.5in,angle=270]{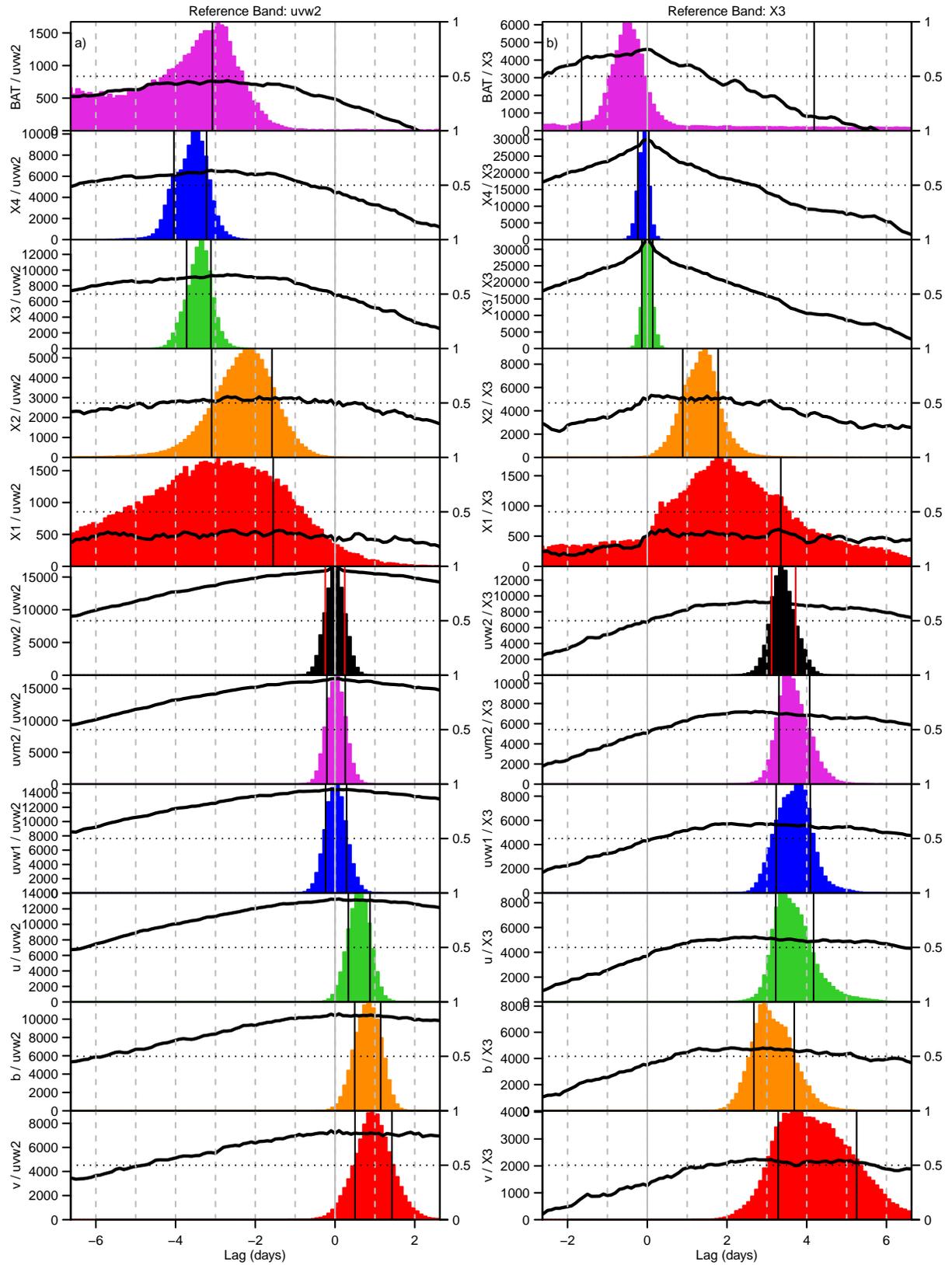} 
\caption{a: Cross-correlation functions (in black; scale on the right) and centroid distributions (in the same color as the light curves in Figure~2; frequency scale on the left) for each band relative to {\it uvw2} as the ``reference band.''
The dotted horizontal line shows $ r = 0.5 $.
Vertical black lines (red for {\it uvw2}) indicate the bounds of the 68\% ($\pm 1 \sigma$) confidence intervals.
The top three panels show strong correlations within the hard X-rays with no measurable interband lag, the fourth panel indicates a significant but weaker correlation with a lag of $\sim$1.5 days, the fifth panel shows essentially zero correlation (so lags are meaningless), and the bottom six panels show that the UV/optical lags behind the X-rays on an $\sim$3 day timescale.
\edit{Note that for BAT and X1 the errors given in Table~6 are so large that the centroid distribution histograms extend outside the figure panel boundaries.}
Figure 4(b): same as Figure 4(a) except for X3 as the reference band.
The UV/optical data are all strongly correlated with no measurable lag above upper limits of $< \pm 0.5$~day within the UV, and an apparent lag of $\lesssim$1~day between {\it uvw2} and the optical bands.}
\label{fig:fig4}
\end{center}
\end{figure}

\begin{figure}
\figurenum{5}
\begin{center}
 \includegraphics[width=3.5in,angle=270]{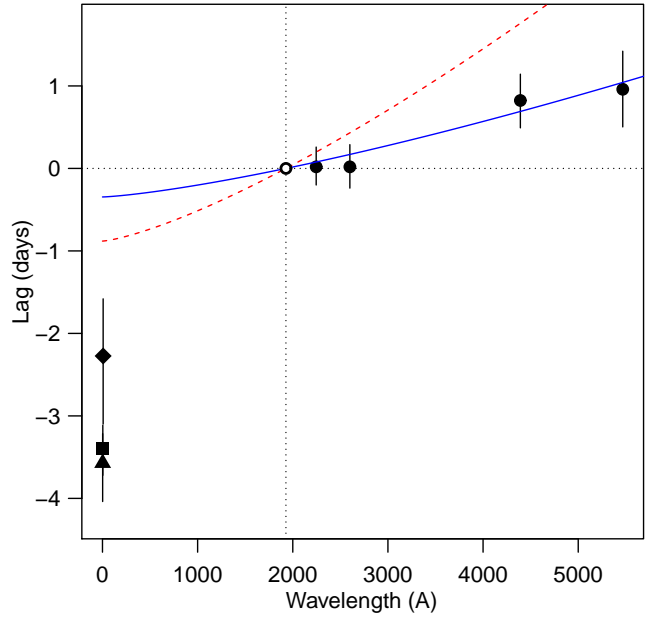} 
\caption{Lag-wavelength fit for the {\it uvw2}-referenced CCFs.
The X4, X3, and X2 lags are shown as a triangle, square, and diamond (respectively), while the UVOT lags are all shown as circles.
Error bars are $\pm 1 \sigma$.
The {\it uvw2} autocorrelation function is shown as an empty circle because the fits are forced to go through that point and it does not participate.
A fit of the function $ \tau = \tau_0 [ (\lambda/\lambda_0)^{4/3} - 1 ] $ to the full participating seven-point dataset is shown as the dashed red line and a fit of the same function to just the four UVOT points is shown as the solid blue line.
While the UV/optical data produce an acceptable fit, that cannot be done if the X-ray data are included.
This is contrary to the expectations of the standard reprocessing model, which predicts that all points should be fit by a $ \tau \propto \lambda^{4/3} $ functional form.}
\label{fig:fig5}
\end{center}
\end{figure}

\begin{figure}
\figurenum{6}
\begin{center}
 \includegraphics[width=3.5in]{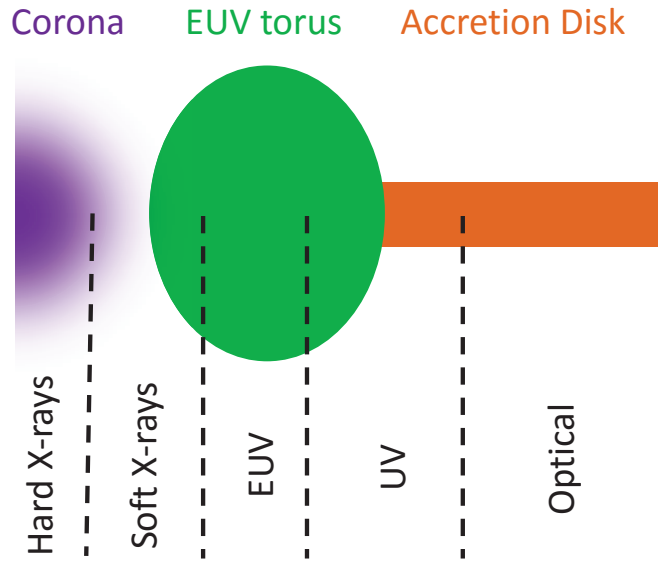} 
\caption{Sketch \edit{(not to scale)} of the proposed geometry of the corona/EUV torus/accretion disk region, adapted from \cite{Gardner17}.
The key difference between this and the standard lamp-post reprocessing model is the addition of the EUV toroidal-shaped component, which prevents the X-ray corona from directly illuminating the accretion disk.
Instead, the corona illuminates and heats the proposed EUV component, which then thermalizes and illuminates the disk.}
\label{fig:fig6}
\end{center}
\end{figure}

\end{document}